# Empirical Analysis of Predictive Algorithms for Collaborative Filtering

**John S. Breese   David Heckerman   Carl Kadie**
Microsoft Research
Redmond, WA 98052-6399
*{breese,heckerma,carlk}@microsoft.com*

## Abstract

Collaborative filtering or recommender systems use a database about user preferences to predict additional topics or products a new user might like. In this paper we describe several algorithms designed for this task, including techniques based on correlation coefficients, vector-based similarity calculations, and statistical Bayesian methods. We compare the predictive accuracy of the various methods in a set of representative problem domains. We use two basic classes of evaluation metrics. The first characterizes accuracy over a set of individual predictions in terms of average absolute deviation. The second estimates the utility of a ranked list of suggested items. This metric uses an estimate of the probability that a user will see a recommendation in an ordered list.

Experiments were run for datasets associated with 3 application areas, 4 experimental protocols, and the 2 evaluation metrics for the various algorithms. Results indicate that for a wide range of conditions, Bayesian networks with decision trees at each node and correlation methods outperform Bayesian-clustering and vector-similarity methods. Between correlation and Bayesian networks, the preferred method depends on the nature of the dataset, nature of the application (ranked versus one-by-one presentation), and the availability of votes with which to make predictions. Other considerations include the size of database, speed of predictions, and learning time.

## 1 Introduction

Typically, automated search over a corpus of items is based on a query identifying intrinsic features of the items sought. Search for textual documents (e.g. Web pages) uses queries containing words or describing concepts that are desired in the returned documents. Search for titles of compact discs, for example, requires identification of desired artist, genre, or time period. Most content retrieval methodologies use some type of similarity score to match a query describing the content with the individual titles or items, and then present the user with a ranked list of suggestions.

A complementary method of identifying potentially interesting content uses data on the preferences of a set of users. Typically, these systems do not use any information regarding the actual content (e.g. words, author, description) of the items, but are rather based on usage or preference patterns of other users. So called collaborative filtering or recommender systems [Resnick and Varian, 1997] are built on the assumption that a good way to find interesting content is to find other people who have similar interests, and then recommend titles that those similar users like.

Though there is increasing commercial interest in collaborative filtering technology, there has been little published research on the relative performance of various algorithms used in collaborative filtering systems. In this paper we describe various collaborative filtering prediction methodologies, including previously published algorithms based on correlation coefficients, as well as algorithms based on learning Bayesian models. We present empirical data regarding the relative predictive performance of the various algorithms and extensions. Although we present some results addressing the computational and scalability issues involved in applying the various algorithms, our primary emphasis is the accuracy and the quality of recommendations of the predictive component.

## 2    Collaborative Filtering Algorithms

The task in collaborative filtering is to predict the utility of items to a particular user (the active user) based on a database of user votes from a sample or population of other users (the user database). In this paper we will examine two general classes of collaborative filtering algorithms. *Memory-based* algorithms operate over the entire user database to make predictions. In *Model-based* collaborative filtering, in contrast, uses the user database to estimate or learn a model, which is then used for predictions.

Collaborative filtering systems are often distinguished by whether they operate over implicit versus explicit votes. Explicit voting refers to a user consciously expressing his or her preference for a title, usually on a discrete numerical scale. For example, GroupLens system of Resnick et al. [1994] uses a scale of one (bad) to five (good) for users to rate Netnews articles, and users explicitly rate each article after reading it. Implicit voting refers to interpreting user behavior or selections to impute a vote or preference. Implicit votes can based on browsing data (for example in Web applications), purchase history (for example in online or traditional stores), or other types of information access patterns.

Regardless of the type of vote data available, collaborative filtering algorithms must address the issue of missing data— we typically do not have a complete set of votes across all titles. We cannot assume that items are missing at random. In most applications, users will vote on items they have accessed, and are more likely to access (and vote) on items they like.

Many of the applications of interest to us involve implicit voting, and some of the algorithms described in the next section rely on an interpretation that any vote appearing in the database indicates a positive preference. We also show that by making different assumptions about the nature of missing data, the performance of collaborative filtering algorithms can be improved.

### 2.1    Memory-Based Algorithms

Generally, the task in collaborative filtering is to predict the votes of a particular user (we will refer to this user as the active user) from a database of user votes from a sample or population of other users. The user database therefore consists of a set of votes $v_{i,j}$ corresponding to the vote for user $i$ on item $j$. If $I_i$ is the set of items on which user $i$ has voted, then we can define the mean vote for user $i$ as:

$$\overline{v}_i = \frac{1}{|I_i|} \sum_{j \in I_i} v_{i,j}$$

In memory-based collaborative filtering algorithms, we predict the votes of the active user (indicated with a subscript $a$) based on some partial information regarding the active user and a set of weights calculated from the user database. We assume that the predicted vote of the active user for item $j$, $p_{a,j}$, is a weighted sum of the votes of the other users:

$$p_{a,j} = \overline{v}_a + \kappa \sum_{i=1}^{n} w(a,i)(v_{i,j} - \overline{v}_i) \qquad (1)$$

where $n$ is the number of users in the collaborative filtering database with nonzero weights. The weights $w(i,a)$ can reflect distance, correlation, or similarity between each user $i$ and the active user. $\kappa$ is a normalizing factor such that the absolute values of the weights sum to unity. In the following, we distinguish between the various collaborative filtering algorithms in terms of the details of the "weight" calculation. There are other possible characterizations for memory-based collaborative filtering, however in this paper we restrict ourselves to the formulation described above.

#### 2.1.1    Correlation

This general formulation of statistical collaborative filtering (as opposed to verbal or qualitative annotations) first appeared in the published literature in the context of the GroupLens project, where the Pearson correlation coefficient was defined as the basis for the weights [Resnick et al., 1994]. The correlation between users $a$ and $i$ is:

$$w(a,i) = \frac{\sum_j (v_{a,j} - \overline{v}_a)(v_{i,j} - \overline{v}_i)}{\sqrt{\sum_j (v_{a,j} - \overline{v}_a)^2 \sum_j (v_{i,j} - \overline{v}_i)^2}} \qquad (2)$$

where the summations over $j$ are over the items for which both users $a$ and $i$ have recorded votes.

#### 2.1.2    Vector Similarity

In the field of information retrieval, the similarity between two documents is often measured by treating each document as a vector of word frequencies and computing the cosine of the angle formed by the two frequency vectors [Salton and McGill, 1983]. We can adopt this formalism to collaborative filtering, where users take the role of documents, titles take the role of words, and votes take the role of word frequencies. Note that under this algorithm, observed votes indicate a positive preference, there is no role for negative votes, and unobserved items receive a zero vote. The relevant weights are now

$$w(a, i) = \sum_j \frac{v_{a,j}}{\sqrt{\sum_{k \in I_a} v_{a,k}^2}} \frac{v_{i,j}}{\sqrt{\sum_{k \in I_i} v_{i,k}^2}} \quad (3)$$

where the squared terms in the denominator serve to normalize votes so that users that vote on more titles will not *a priori* be more similar to other users. Other normalization schemes, including absolute sum and number of votes, are possible.

## 2.2  Extensions to Memory-Based Algorithms

We have investigated a number of modifications to the standard algorithms that can improve performance. We describe these extensions here and the effectiveness of each is discussed in Section 4.

### 2.2.1  Default Voting

Default voting is an extension to the correlation algorithm described in Section 2.1.1. It arose out of the observation that when there are relatively few votes, for either the active user or the matching user, the correlation algorithm will not do well because it uses only votes in the intersection of the items both individuals have voted on ($I_a \cap I_j$). If we assume some default value as a vote for titles for which we do not have explicit votes, then we can form the match over the union of voted items,($I_a \cup I_j$), where the default vote value is inserted into the formula for the appropriate unobserved items.

In addition, we can assume the same default vote value $d$ for some number of additional items $k$ that neither user has voted on. This has the effect of assuming there are some additional number of unspecified items that neither user voted on, but they would nonetheless agree on.[1] In most cases, the value for $d$ will reflect a neutral or somewhat negative preference for these unobserved items.

In applications with implicit voting, an observed vote is typically an indication of a positive preference (e.g. a visit to the Web page is assigned a vote value of 1). In this case the default vote can take on the value associated with "did not visit" or 0. In this instance, default voting takes on the role of extending the data for each user with the true value for missing data. Note, however, we only calculate weights for users who match the active user on at least one item.

---

[1]In our experiments, we have used a value of 10,000 or $k$.

### 2.2.2  Inverse User Frequency

In applications of vector similarity in information retrieval, word frequencies are typically modified by the *inverse document frequency* [Salton and McGill, 1983]. The idea is to reduce weights for commonly occurring words, capturing the intuition that they are not as useful in identifying the topic of a document, while words that occur less frequently are more indicative of topic. We can apply an analogous transformation to votes in a collaborative filtering database, which we term *inverse user frequency*. The idea is that universally liked items are not as useful in capturing similarity as less common items. We define the $f_j$ as $\log \frac{n}{n_j}$ where $n_j$ is the number of users who have voted for item $j$ and $n$ is the total number of users in the database. Note that if everyone has voted on a item $j$, then the $f_j$ is zero.

To apply inverse user frequency while using the vector similarity algorithm, we use a transformed vote in Equation 3. The transformed vote is simply the original vote multiplied by the $f_j$ factor. In the case of correlation, we modify Equation 2 so that the $f_j$ is treated as a frequency and an item with a higher $f_j$ is assigned more weight in the correlation calculation. The relevant correlation weight with inverse frequency is:

$$w(a, i) = \frac{\sum_j f_j \sum_j f_j v_{a,j} v_{i,j} - (\sum_j f_j v_{a,j})(\sum_j f_j v_{i,j})}{\sqrt{UV}}$$

where

$$U = \sum_j f_j (\sum_j f_j v_{a,j}^2 - (\sum_j f_j v_{a,j})^2)$$

$$V = \sum_j f_j (\sum_j f_j v_{i,j}^2 - (\sum_j f_j v_{i,j})^2)$$

### 2.2.3  Case Amplification

Case amplification refers to a transform applied to the weights used in the basic collaborative filtering prediction formula as in Equation 1. We transform the estimated weights as follows

$$w'_{a,i} = \begin{cases} w_{a,i}^\rho & \text{if } w_{a,i} \geq 0 \\ -(-w_{a,i}^\rho) & \text{if } w_{a,i} < 0 \end{cases}$$

The transform emphasizes weights that are closer to one, and punishes low weights. A typical value for $\rho$ for our experiments is 2.5.

## 2.3   Model-Based Methods

From a probabilistic perspective, the collaborative filtering task can be viewed as calculating the expected value of a vote, given what we know about the user. For the active user, we wish to predict votes on as-yet unobserved items. If we assume that the votes are integer valued with a range for 0 to $m$ we have:

$$p_{a,j} = E(v_{a,j}) = \sum_{i=0}^{m} \Pr(v_{a,j} = i | v_{a,k}, k \in I_a)\, i \quad (4)$$

where the probability expression is the probability that the active user will have a particular vote value for item $j$ given the previously observed votes. In this paper we examine two alternative probabilistic models for collaborative filtering, cluster models and Bayesian networks.

### 2.3.1   Cluster Models

One plausible probabilistic model for collaborative filtering is a Bayesian classifier where the probability of votes are conditionally independent given membership in an unobserved class variable $C$ taking on some relatively small number of discrete values. The idea is that there are certain groups or types of users capturing a common set of preferences and tastes. Given the class, the preferences regarding the various items (expressed as votes) are independent. The probability model relating joint probability of class and votes to a tractable set of conditional and marginal distributions is the standard "naive" Bayes formulation:

$$\Pr(C = c, v_1, \ldots, v_n) = \Pr(C = c) \prod_{i=1}^{n} \Pr(v_i | C = c)$$

The left-hand side of this expression is the probability of observing an individual of a particular class and a complete set of vote values. It is straightforward to calculate the needed probability expressions for Equation 4 within this framework. This model is also known as a multinomial mixture model.

The parameters of the model, the probabilities of class membership $\Pr(C = c)$, and the conditional probabilities of votes given class $\Pr(v_i | C = c)$ are estimated from a training set of user votes, the user database. Since we never observe the class variables in the database of users, we must employ methods that can learn parameters for models with hidden variables. We use the EM algorithm [Dempster et al., 1977] to learn the parameters for a model structure with a fixed number of classes. We choose the number of classes by selecting the model structure that yields the largest (approximate) marginal likelihood of the data. We use the method of Cheeseman and Stutz (1995) to approximate the marginal likelihood (see also Chickering and Heckerman, 1997). In our experiments, we assume each model structure (every possible number of classes) is equally likely, and use a uniform prior for model parameters. We initialize the EM algorithm using the marginal-plus-noise technique described in [Thiesson et al., 1997].

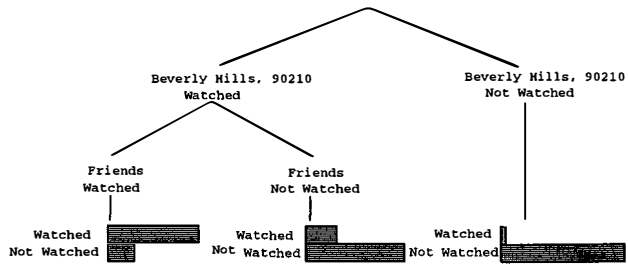

Figure 1: A decision tree for whether an individual watched "Melrose Place", with parents "Friend's", and "Beverly Hills, 90201". The bar charts at the bottom of the tree indicate the probabilities of watched and not watched for "Melrose Place", conditioned on viewing the parent programs.

### 2.3.2   Bayesian Network Model

An alternative model formulation for probabilistic collaborative filtering is a Bayesian network with a node corresponding to each item in the domain. The states of each node correspond to the possible vote values for each item. We also include a state corresponding to "no vote" for those domains where there is no natural interpretation for missing data.

We then apply an algorithm for learning Bayesian networks to the training data, where missing votes in the training data are indicated by the "no vote" value. The learning algorithm searches over various model structures in terms of dependencies for each item. In the resulting network, each item will have a set of parent items that are the best predictors of its votes. Each conditional probability table is represented by a decision tree encoding the conditional probabilities for that node. An example of such a tree, for television viewing data (see Section 3.2) is shown in Figure 1. Details of the learning algorithm are discussed in Chickering et al.(1997). In the remainder of the paper the term Bayesian network will refer to these networks with a decision tree for each title.

In the experiments that follow, we use a *structure prior* that penalizes each additional free parameter with probability 0.1, and derive parameter priors from a prior network as described in Chickering et al., 1997.

In particular, we use a prior network that encodes a uniform distribution over all possible outcomes and an equivalent sample size of 10. Experiments on subsets of the training data showed these parameters to produce accurate results, although there was little sensitivity.

## 3  Empirical Analysis

The purpose of this paper is to evaluate the predictive accuracy of the various algorithms for collaborative filtering. In this section we will describe the evaluation criteria, the various protocols, and the datasets used in the analysis. We then present and discuss the results regarding predictive accuracy, as well as several computational considerations.

### 3.1  Evaluation Criteria

The effectiveness of a collaborative filtering algorithm depends on manner in which recommendations will be presented to the user. To evaluate these algorithms, we have defined metrics based on the type of collaborative filtering application and interface one is providing.

There are two basic classes of collaborative filtering applications. In the first class, individual items are presented one-at-a-time to the users along with a rating indicating potential interest in the topic. The original GroupLens system was in this category— each article in a GNUs-like Netnews interface has an ASCII barchart indicating the system's prediction regarding the user's possible interest in that article. Thus, each piece of content has an associated estimated rating, and the user interface displays this estimate along with a link to the content or as a part of the display or presentation of the item.

A second class of collaborative filtering applications present the user with an ordered list of recommended items. Examples of systems that present recommendation lists include PHOAKS [L.Terveen et al., 1997] and SiteSeer [Rucker and Polanco, 1997]. In the spirit of the Internet search engines, these systems provide a ranked list of items (Web sites, music recordings) where highest ranked items are predicted to be most preferred. In these types of systems, the user presumably will investigate items in the ordered list starting at the top hoping to find interesting items.

We have applied two scoring metrics in our evaluations–one appropriate for individual item-by-item recommendations and the other appropriate for ranked lists. In both cases, the basic evaluation sequence proceeds as follows. A dataset of users (and their votes) is divided into a training set and a test set. The data for the training set is used as the col-laborative filtering database or to build a probabilistic model. We then cycle through the users in the test set, treating each user as the *active* user. We divide the votes for each test user into a set of votes that we treat as observed, $I_a$, and a set that we will attempt to predict, $P_a$. We use the votes in $I_a$ to predict the votes in $P_a$ as shown in Equations 1 and 4.

For individual scoring, we look at the average absolute deviation of the predicted vote to the actual vote on items the users in the test set have actually voted on. That is, if the number of predicted votes in the test set for the active case is $m_a$, then the average absolute deviation for a user is:

$$S_a = \frac{1}{m_a} \sum_{j \in P_a} |p_{a,j} - v_{a,j}|$$

These scores are then averaged over all the users in the test set of users. This metric was also used in evaluating the GroupLens project [Miller et al., 1997].

For ranked scoring, the story is a bit more complex. In information retrieval research, ranked lists of returned items are evaluated in terms of recall and precision. For a given number of returned items, recall is the percentage of relevant items that were returned and precision is the percentage of returned items that are relevant. In a collaborative filtering framework, if votes were binary (like and dislike) and we had complete preference judgments for a set of users we could develop a similar metric. However, more generally, we wish to estimate the expected utility of a particular ranked list to a user. The expected utility of a list is simply the probability of viewing a recommended item times its utility. In this analysis, we will equate the utility of an item with the difference between the vote and the default or neutral vote in the domain.

Furthermore, we make an estimate of how likely it is that the user will visit an item on a ranked list. We posit that each successive item in a list is less likely to be viewed by the user with an exponential decay. Then the expected utility of a ranked list of items (sorted by index $j$ in order of declining $v_{a,j}$) is:

$$R_a = \sum_j \frac{\max(v_{a,j} - d, 0)}{2^{(j-1)/(\alpha-1)}} \qquad (5)$$

where $d$ is the neutral vote and $\alpha$ is the viewing halflife. The halflife is the number of the item on the list such that there is a 50-50 chance the user will review that item. For these experiments, we used a halflife of 5 items. [2]

---

[2] We ran a set of experiments using a halflife of 10 items and found little sensitivity of results.

In scoring a ranked list generated for a user, we apply Equation 5 using observed votes where available. For items that are not available, we apply the neutral vote, $d$, which effectively removes those items from the scoring. The final score for an experiment over a set of active users in the test set is

$$R = 100 \frac{\sum_a R_a}{\sum_a R_a^{max}}$$

where $R_a^{max}$ is the maximum achievable utility if all observed items had been at the top of the ranked list, ordered by vote value. This transformation allows us to consider results independent of the size of the test set and number of items predicted in a given experiment.

### 3.2 Datasets

We evaluated the algorithm for three separate datasets, as follows:

- **MS Web**: This dataset captures individual visits to various areas (vroots) of the Microsoft corporate web site. This is an example of an implicit voting database and application. Each vroot was characterized as being visited (vote of one) or not (no vote).

- **Television**: This dataset uses Neilsen network television viewing data for individuals for a two week period in the summer of 1996. The data was transformed into binary form indicating whether each show was watched, or not, as above.[3]

- **EachMovie**: This is an explicit voting example using data from the EachMovie collaborative filtering site deployed by Digital Equipment Research Center from 1995 through 1997. [4] Votes ranged in value from 0 to 5.

Table 3.2 provides additional information about each dataset.

### 3.3 Protocols

We did two classes of experiments reflecting differing numbers of votes available to the recommenders. In the first protocol, we withhold a single randomly selected vote for each user in the test set, and try to predict its value given all the other votes the user has voted on. We term this protocol *All but 1*. In the second set of experiments, we randomly select 2, 5, or 10 votes from each test user as the observed votes, and then attempt to predict the remaining votes. We call these protocols *Given 2*, *Given 5*, and *Given 10*.

---

[3]This dataset was made available for this study courtesy of Nielsen Media Research.
[4]For more information see *http://www.research.digital.com/SRC/EachMovie/*.

|              | Dataset |         |           |
|--------------|---------|---------|-----------|
|              | MSWEB   | Neilsen | Eachmovie |
| Total users  | 3453    | 1463    | 4119      |
| Total titles | 294     | 203     | 1623      |
| Mean votes per user | 3.95 | 9.55 | 46.4 |
| Median votes per user | 3 | 8 | 26 |

Table 1: Number of users, titles, and votes for the datasets used in testing the algorithms. Only users with 2 or more votes are considered.

The *All but 1* experiments measure the algorithms' performance when given as much data as possible from each test user. The various *Given* experiments look at users with less data available, and examine the performance of the algorithms when there is relatively little known about an active user. In running the tests, if a prospective test did not have adequate votes for a trial it was eliminated from the evaluation. Thus the number of trials evaluated under each protocol vary.

## 4  Results

In the following sections, we compare algorithms and analyze the effects of individual algorithmic extensions. We use randomized block design where each algorithm is run on the same test cases and observed votes. We will refer to one of these comparisons as an experiment. Our analyses uses ANOVA with the Bonferroni procedure for multiple comparisons statistics [McClave and Dietrich, 1988]. In the tables that follow, the value in the last row is labeled *RD* for *Required Difference*. The difference between any two scores in a column must be at least as big as the value in the *RD* row in order to be considered statistically significant at the 90% confidence level for the experiment as a whole. As a visual aid, a score in **boldface** is significantly different from the score directly below it in the table.

### 4.1  Overall Performance

The following tables show the performance of the various major classes of algorithms on the various datasets and experiments. We compared the best performing variation of each algorithm on each dataset, for the different protocols. We also present the scores that result from presenting the user with the most popular items, regardless of the known votes of the individ-ual. This results in a baseline performance of a "zero-order" collaborative filtering system, and is labeled as POP in the tables. The algorithm labeled CR+ refers to use of the correlation technique with inverse user frequency, default voting, and case amplification extensions. VSIM refers to using the vector similarity method with the inverse user frequency transformation. BN and BC refer to the Bayesian network and clustering models respectively.

| | MS Web, Rank Scoring | | | |
|---|---|---|---|---|
| Algorithm | Given2 | Given5 | Given10 | AllBut1 |
| BN | 59.95 | **59.84** | 53.92 | **66.69** |
| CR+ | **60.64** | 57.89 | 51.47 | 63.59 |
| VSIM | **59.22** | 56.13 | 49.33 | **61.70** |
| BC | **57.03** | **54.83** | **47.83** | **59.42** |
| POP | 49.14 | 46.91 | 41.14 | 49.77 |
| *RD* | *0.91* | *1.82* | *4.49* | *0.93* |

Table 2: Ranked scoring results for the MS Web dataset. Higher scores indicate better performance.

| | Neilsen, Rank Scoring | | | |
|---|---|---|---|---|
| Algorithm | Given2 | Given5 | Given10 | AllBut1 |
| BN | 34.90 | 42.24 | **47.39** | **44.92** |
| CR+ | **39.44** | **43.23** | **43.47** | **39.49** |
| VSIM | 39.20 | **40.89** | **39.12** | 36.23 |
| BC | **19.55** | 18.85 | **22.51** | **16.48** |
| POP | 20.17 | 19.53 | 19.04 | 13.91 |
| *RD* | *1.53* | *1.78* | *2.42* | *2.40* |

Table 3: Ranked scoring results for the Neilsen dataset. Higher scores indicate better performance.

| | EachMovie, Rank Scoring | | | |
|---|---|---|---|---|
| Algorithm | Given2 | Given5 | Given10 | AllBut1 |
| CR+ | **41.60** | **42.33** | **41.46** | **23.16** |
| VSIM | **42.45** | **42.12** | **40.15** | 22.07 |
| BC | **38.06** | **36.68** | **34.98** | **21.38** |
| BN | 28.64 | 30.50 | 33.16 | **23.49** |
| POP | 30.80 | 28.90 | 28.01 | 13.94 |
| *RD* | *0.75* | *0.75* | *0.78* | *0.78* |

Table 4: Ranked scoring results for the EachMovie dataset. Higher scores indicate better performance.

Our results show that Bayesian networks with decision trees at each node and correlation methods are the best performing algorithms over the experiments we have run. We ran 16 combinations of dataset, protocol, and scoring criteria. The Bayesian network and correlation-based were each either best, or statistically equivalent, in 10 cases. Bayesian clustering was best performing in 2 cases and vector similarity was best in 3 cases.

We see that the Bayesian network performs best under the *All but 1* protocol. Generally, all the methods perform less well in the *Given 2* and *Given 5* protocols as might be expected. However the vector similarity and clustering methods are competitive for some of these limited-data scenarios, since these methods can use partial information effectively.

Table 2 shows data for rank scoring for the Microsoft web site dataset. For ranked scoring, higher scores indicate better performance. We see the Bayesian network model results in the best, or statistically equivalent to the best, score for all protocols. Correlation, with the appropriate enhancements designed to improve ranked scoring, is fairly close in performance. Note that correlation without default voting cannot operate on binary data with implicit voting, since all observed votes will have the same value. The vector similarity algorithm is slightly worse than correlation. All these algorithms outperform using popularity as a recommender.

For the Neilsen dataset (Table 3), the Bayesian network outperforms the other algorithms except for the

*Given 2* protocol. Correlation, with extensions, and vector similarity are fairly close in performance, while Bayesian clustering performs relatively poorly. We see that the Bayesian network drops off in performance quite significantly for the *Given 2* protocol, relative to correlation and vector similarity. We will discuss this observation below.

We see a somewhat different pattern for EachMovie under ranked scoring, shown in Table 4. Here the correlation algorithm is the top performer overall, with vector similarity performing well with less data. For this dataset and score, the Bayesian network performs worse than any of the other algorithms on all the *Given* experiments, but is the top performer and is competitive with correlation for the *All but 1* protocol.

The Bayesian networks using decision trees suffer in the *Given* scenarios because they are provided with relatively little data. If a title that is held out for testing appears near the top of a tree, then it's value is set to "no vote" in evaluating the probability of a possibly related title. This may result in a title that *is* provided being ignored or having little impact, simply due to the ordering of the various predicting titles in the tree. The various *All But 1* experiments are able to utilize trees to a fuller extent, and therefore perform well relative to the other methods that can use partial data.

| | EachMovie, Absolute Deviation | | | |
|---|---|---|---|---|
| Algorithm | Given2 | Given5 | Given10 | AllBut1 |
| CR | **1.257** | 1.139 | **1.069** | **0.994** |
| BC | 1.127 | 1.144 | 1.138 | 1.103 |
| BN | **1.143** | **1.154** | **1.139** | **1.066** |
| VSIM | **2.113** | **2.177** | **2.235** | **2.136** |
| *RD* | *0.022* | *0.023* | *0.025* | *0.043* |

Table 5: Absolute Deviation scoring results for the EachMovie dataset. Lower scores are better.

For absolute deviation, we examined the EachMovie dataset and results are shown in Table 5. This dataset has a vote range of 0 to 5, making vote prediction a relevant task. We examine the same algorithms as in the previous table, except now we use a correlation algorithm without applying any of the extensions except for inverse user frequency. The other extensions are not effective for absolute deviation scoring. This basic correlation algorithm performs best in all but the *Given 2* experiments, indicating that this algorithm performs well when given adequate data regarding the active case. The Bayesian clustered model does slightly better than the Bayesian network, and outperforms correlation in the *Given 2* and *Given 5* cases.

### 4.2    Inverse User Frequency

In Section 2.2.2 we describe using inverse user frequency to modify vote values in applying memory-based algorithms. We performed a set of 12 experiments (3 datasets, 4 protocols) each for vector similarity and correlation judging the effect of applying inverse user frequency under ranked scoring. In all experiments, application of IUF improved the ranked score, and in 23 of 24 cases results were statistically significant. The average improvement was 1.9%, with an improvement of 2.2% for the vector similarity algorithm, and 1.5% for the correlation algorithm.

In 8 experiments run on the EachMovie dataset using absolute deviation scoring, the improvement averaged a more impressive 11%. Results were significant in 6 of the 8 experiments. The average improvement was of 15.5% for vector similarity, and 6.5% for correlation.

### 4.3    Case Amplification

Case amplification (Section 2.2.3) modifies weights used in a memory-based algorithm to emphasize higher weights. We performed a set of 12 experiments (3 datasets, 4 protocols) applying case amplification to correlation. The average improvement in the ranked score was 4.8%, and results were significant in 11 of 12 experiments. There is no significant effect of case amplification on absolute deviation scoring. We also ran experiments combining case amplification and inverse user frequency, and found the benefits to be additive.

### 4.4    Probabilistic Methods

We used a training set to build probabilistic models for each dataset. Each title was encoded with an additional explicit vote value of "no vote" to complete the dataset for probabilistic learning. When scoring with Bayesian networks and cluster models, the "no vote" values were explicitly entered into the network when missing, for both ranked and absolute deviation scoring. For the trees, the "no vote" values were entered in each tree independently in order to generate a probability for that title. For absolute deviation scoring, the expected vote was calculated by renormalizing the output probabilities, clamping the "no vote" probability to zero.

There are roughly 1600 movies in the EachMovie dataset, too many to estimate a full model in a reasonable amount of time. Therefore the Bayesian methods were trained from EachMovie for the top 300 movies in terms of overall popularity. For testing, all 1600 movies were used. In the other datasets, all items were used for training and testing.

For the Bayesian networks, we applied alternate prior specifications which resulted in trees of varying complexity. Priors that strongly penalized splits generated Bayesian networks with nodes with approximately 2 to 4 parents and 4 to 6 distributions in the decision tree representation. The model with the larger trees had somewhere between 4 and 6 predecessors and 6 to 8 distributions per variable. In all our experiments the larger trees outperformed the smaller tree so we restrict our results to those models. Additional details are available in Breese *et al.* (1998).

Applying clustering to the datasets identified 3 classes for the Neilsen dataset, 7 classes for the MS Web dataset, and 9 classes for the EachMovie dataset. The classes found by clustering for the MS Web dataset are shown below. Each entry is a page area or virtual root that distinguishes this class from the others. The class names on the left were manually generated based on inspecting the resulting classes.

**Support**  Support Desktop, Knowledge Base, Windows95 Support, Search, NT Server Support

**Windows**  Products, Free Downloads, Windows95, Windows95 Support, Windows Family of Products

**Office**  Products, MS Office Info, Free Downloads, MS

Word News, Office Free Stuff, MS Office

**Developers** Search, Training, Games, Developer Network, Job Openings

**Internet Explorer** Internet Explorer, Free Downloads, IE support, Net Meeting, International IE Content

**Internet Explorer Technical** Search, Free Downloads, Products, Internet Explorer, Internet Site Construction for Dev.

**IE Site Builder** Internet Site Construction for Dev., Web Site Builders Gallery, Developer Workshop, Sitebuilder Network Membership, Jakarta, ActiveX Technology Dev.

Among probabilistic methods, the Bayesian network with a decision tree at each item outperformed the cluster models for ranked scoring. In 12 comparisons, there was an average 41% improvement in ranked scores, all differences being statistically significant. For absolute deviation experiments run with the EachMovie data, we found that the cluster model performed slightly better than the trees.

## 5 Additional Issues

Although predictive accuracy is probably the most important aspect in gauging the efficacy of a collaborative filtering algorithm, there are other considerations, including size of model, sampling, and runtime performance.

If one considers the size of the overall collaborative filtering prediction representation, memory-based methods require a relatively small algorithm code base, plus a user database consisting of a sample of user votes. The model-based methods require the representation of the Bayesian network model, typically having much smaller memory requirements. For example, the user databases required for the memory-based methods for the EachMovie and MS Web datasets were approximately 314 and 318 Kilobytes compressed, while the Bayesian network model sizes were 27 and 55 Kilobytes compressed respectively.

The number of items in the usage database used for the memory-based methods was determined by experimenting with the scoring for various sizes of training set. Figure 2 shows the increase in ranked scoring accuracy as a function of size of training set. We used training set sizes (number of users) of 1637 for Neilsen, 5000 for EachMovie, and 32711 for MS Web. Identical training sets were used as the user database for model-based methods, and as the database for learning probabilistic models. Our experiments have found

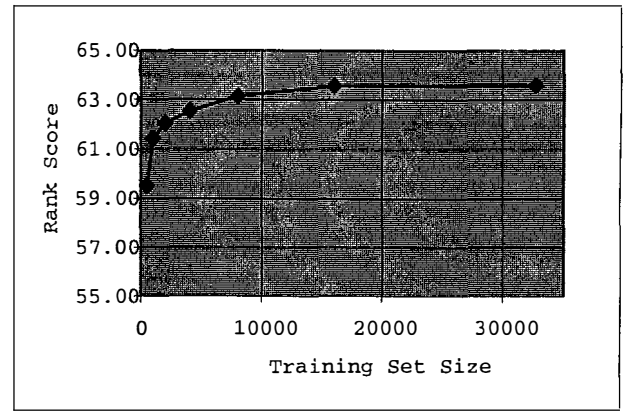

Figure 2: A learning curve showing the effect of sample size on ranked scoring for the correlation method, *All but 1* protocol, MSWeb dataset.

that sample sizes on this order are adequate for purposes of generating recommendations.

In terms of runtime performance, the probabilistic, model-based methods were approximately 4 times as fast as the memory-based methods in generating recommendations, with correlation generating 3.2 recommendations per second and the Bayes net generating 12.9 recommendations per second on 266 MHz Pentium II processor (Eachmovie dataset). Of course, the probabilistic models must be learned. Learning times for the models used in these experiments ranged from less than an hour for Neilsen and up to 8 hours for EachMovie and MS Web.

## 6 Conclusions

This paper presents an extensive set of experiments regarding the predictive performance of statistical algorithms for collaborative filtering or recommender systems. Results indicate that for a wide range of conditions, Bayesian networks with decision trees at each node and correlation methods outperform Bayesian-clustering and vector similarity methods. Between correlation and Bayesian networks, the preferred method depends on the nature of the dataset, nature of the application (ranked or one-by-one presentation), and the availability of votes with which to make predictions. We see that when there are relatively few votes, correlation and Bayesian networks have less of an advantage over the other techniques.

Other considerations include the size of database, speed of predictions, and learning time. Bayesian networks are typically have smaller memory requirements and allow for faster predictions than a memory-based technique such as correlation. However, the Bayesian methods examined here require a learning phase that can take up to several hours and results in a lag before changed behavior is reflected in recommendations.

We plan to make the MS Web data used in this study available to learning community through the Irvine repository. As noted, the EachMovie data is currently available. We hope that the availability of this data coupled with discussion spurred by this paper will result in additional examination and improvement of collaborative filtering algorithms.

**Acknowledgements**

Datasets for this paper were generously provided by Digital Equipment Corporation (EachMovie), Neilsen Media Research (Neilsen), and Microsoft Corporation (MS Web). Max Chickering, David Hovel, and Robert Rounthwaite contributed to the programming of the algorithms that were used in this study. We also thank Max Chickering, Eric Horvitz, and Chris Meek for useful discussions. John Riedl also provided useful comments.